\documentclass[aps,prl,groupedaddress,fleqn,twocolumn]{revtex4}
\pdfoutput=1
\usepackage{graphicx}
\usepackage{amsmath}
\usepackage{gensymb}
\begin{document}

\title{
Lagrangian formulation and symmetrical description of liquid dynamics
%Lagrangian formulation of liquid dynamics and equivalence of hydrodynamic and solid-like elastic description
}
\author{K. Trachenko$^{1}$}
\address{$^1$ School of Physics and Astronomy, Queen Mary University of London, Mile End Road, London, E1 4NS, UK}

\begin{abstract}
Theoretical description of liquids has been primarily based on the hydrodynamic approach and its generalization to the solid-like regime. We show that the same liquid properties can be derived starting from solid-like equations and generalizing them to account for the hydrodynamic flow. Both approaches predict propagating shear waves with the notable gap in $k$-space. This gives an important symmetry of liquids regarding their description. We subsequently construct a two-field Lagrangian of liquid dynamics where the dissipative hydrodynamic and solid-like terms are treated on equal footing. The Lagrangian predicts two gapped waves propagating in opposite space-time directions. The dissipative and mass terms compete by promoting gaps in $k$-space and energy, respectively. When bare mass is close to the field hopping frequency, both gaps close and the dissipative term annihilates the bare mass term in the dispersion relation.
\end{abstract}

%\pacs{61.43.Fs, 64.70.Pf, 61.20.Lc}

\maketitle

The smallness of atomic displacements in solids and weakness of interactions in gases simplify their theoretical description. Liquids do not have those simplifying features (small parameter): they combine large displacements with strong interactions. For this reason, liquids are believed to be not amenable to theoretical understanding at the same level as gases and solids \cite{landau}. The first-principles description of liquids involving the equations of motion for each atom meets a formidable challenge: one deals with coupled non-linear oscillators, the problem not currently tractable due to its exponential complexity \cite{ropp}.

Theoretical description of liquids involves a continuum approximation and, because liquids flow, the hydrodynamic approximation is used \cite{hydro}. For example, the Navier-Stokes equation describes liquid flow and features viscosity as an important flow property. At the same time, liquid properties such as density, bulk modulus and heat capacity are close to those of solids \cite{ropp}. An important solid-like property is the liquid ability to support high-frequency solid-like shear waves. Predicted by Frenkel \cite{frenkel}, this property has been seen in experiments \cite{grim,scarponi,burkel,rec-review,hoso,mon-na,sn} and modeling results, although with a long time lag. Frenkel's idea was that liquid particles oscillate as in solids for some time and then diffusively move to neighbouring quasi-equilibrium positions. He introduced $\tau$ as the average time between diffusive jumps and predicted that liquids behave like solids and hence support shear modes at time shorter than $\tau$, or frequency

\begin{equation}
\omega>\omega_{\rm F}=\frac{1}{\tau}
\label{omegaF}
\end{equation}

Solid-like shear modes are absent in the hydrodynamic description operating when $\omega\tau<1$, whereas (\ref{omegaF}) implies the opposite regime $\omega\tau>1$. To account for the shear modes and other solid-like properties, liquid theories designate the hydrodynamic description as a starting point and subsequently generalize it to account for liquid response at large $\omega$ and wavevector $k$. Several ways of doing so have been proposed, giving rise to a large field of generalized hydrodynamics \cite{boon,baluca,hansen}. This approach was used to describe non-hydrodynamic liquid properties, but faced issues related to its phenomenological character as well as assumptions and extrapolations used (see, e.g. \cite{rec-review,mon-na}).

The traditional hydrodynamic approach to liquids is supported by our common experience that liquids flow and hence necessitate hydrodynamic flow equations as a starting point. This reflects our experience with common low-viscous liquids where $\tau$ is much shorter than observation time. However, flow is less prominent in liquids with large $\tau$ (e.g. liquids approaching glass transition) where properties become more solid-like  \cite{dyre}.

Importantly, on general physical grounds, the picture of combined solid-like oscillatory and diffusive particle motion in liquids gives no preference to a hydrodynamic description of liquids as a starting point. This picture should equally allow for a liquid theory starting from a solid (this is also apparent in the Maxwell interpolation discussed below). There is therefore an important question as to what extent we can use the solid-like approach to liquids as a starting point on par with the hydrodynamic approach. The importance of this question is also highlighted by fairly recent experiments demonstrating solid-like properties of liquids \cite{burkel,rec-review,hoso,mon-na,sn}. This issue could be clarified by examining solid-state and hydrodynamic terms in the liquid Lagrangian, however a Lagrangian formulation of dissipative processes in liquids and gases was deemed impossible (see, e.g., \cite{leut}).

In this paper, we show that important liquid properties can be derived in approaches starting from either hydrodynamic or solid state equations. We consider solid-like transverse modes and their notable gap in $k$-space as an important case study. This suggests that liquids occupy a symmetrical place between the hydrodynamic and solid-like approaches from the physical point of view. This general result also has practical implications for liquid theory. We subsequently construct a two-field Lagrangian accounting for both solid-like and dissipative effects. The Lagrangian predicts two gapped waves propagating in the opposite space-time directions and an interesting interplay between the dissipative and mass terms.

We start with recalling how solid-like transverse modes and their gap in $k$-space can be derived starting from the hydrodynamic approach and generalizing it to endow the liquid with short-time elastic response. This programme involves Maxwell interpolation \cite{maxwell}:

\begin{equation}
\frac{ds}{dt}=\frac{P}{\eta}+\frac{1}{G}\frac{dP}{dt}
\label{a1}
\end{equation}
\noindent where $s$ is shear strain, $\eta$ is viscosity, $G$ is shear modulus and $P$ is shear stress.

(\ref{a1}) reflects Maxwell's statement that shear deformation in a liquid is the sum of the viscous and elastic deformations, given by the first and second right-hand side terms. As mentioned above, both deformations are treated in (\ref{a1}) on equal footing.

Frenkel proposed \cite{frenkel} to represent the Maxwell interpolation by introducing the operator $A$ as

\begin{equation}
A=1+\tau\frac{d}{dt}
\label{a2}
\end{equation}

Then, Eq. (\ref{a1}) can be written in the operator form as

\begin{equation}
\frac{ds}{dt}=\frac{1}{\eta}AP
\label{a3}
\end{equation}

In (\ref{a2})-(\ref{a3}), $\tau$ formally is Maxwell relaxation time $\frac{\eta}{G}$. At the microscopic level, Frenkel's theory approximately identifies this time with the time between consecutive diffusive jumps in the liquid \cite{frenkel}. This is supported by numerous experiments \cite{dyre}.

(\ref{a1}-\ref{a3}) enable us to generalize $\eta$ to account for short-term elasticity and to generalize $G$ to allow for long-time hydrodynamic flow \cite{frenkel}. Indeed, accounting for both long-time viscosity and short-time elasticity in (\ref{a1}) is equivalent to generalizing viscosity as

\begin{equation}
\frac{1}{\eta}\rightarrow\frac{1}{\eta}\left(1+\tau\frac{d}{dt}\right)
\label{sub2}
\end{equation}

$G$ is generalized by noting that if $A^{-1}$ is the reciprocal operator to $A$, (\ref{a3}) can be written as $P=\eta A^{-1}\frac{ds}{dt}$. Because $\frac{d}{dt}=\frac{A-1}{\tau}$ from (\ref{a2}), $P=G(1-A^{-1})s$. Comparing this with the solid-like equation $P=Gs$, we see that the presence of hydrodynamic viscous flow is equivalent to the substitution of $G$ by the operator

\begin{equation}
M=G(1-A^{-1})
\label{operator}
\end{equation}

Adopting the hydrodynamic approach as a starting point of liquid description, we write the Navier-Stokes equation as

\begin{equation}
\nabla^2{\bf v}=\frac{1}{\eta}\left(\rho\frac{d{\bf v}}{dt}+\nabla p\right)
\label{navier2}
\end{equation}

\noindent where ${\bf v}$ is velocity, $\rho$ is density and the full derivative is $\frac{d}{dt}=\frac{\partial}{\partial t}+{\bf v\nabla}$ and generalize $\eta$ according to (\ref{sub2}):

\begin{equation}
\eta\nabla^2{\bf v}=\left(1+\tau\frac{d}{dt}\right)\left(\rho\frac{d{\bf v}}{dt}+\nabla p\right)
\label{gener}
\end{equation}

Having proposed Eq. (\ref{gener}), Frenkel did not analyze it or its solutions. We recently solved Eq. (\ref{gener}) \cite{ropp} and considered the absence of external forces, $p=0$ and the slowly-flowing fluid so that $\frac{d}{dt}=\frac{\partial}{\partial t}$. Then, Eq. (\ref{gener}) reads

\begin{equation}
\eta\frac{\partial^2v}{\partial x^2}=\rho\tau\frac{\partial^2v}{\partial t^2}+\rho\frac{\partial v}{\partial t}
\label{gener2}
\end{equation}

\noindent where $v$ is the velocity component perpendicular to $x$.

In contrast to the Navier-Stokes equation, Eq. (\ref{gener2}) contains the second time derivative of $v$ and hence allows for propagating waves. Using $\eta=G\tau=\rho c^2\tau$, where $c$ is the shear wave velocity, we re-write Eq. (\ref{gener2}) as

\begin{equation}
c^2\frac{\partial^2v}{\partial x^2}=\frac{\partial^2v}{\partial t^2}+\frac{1}{\tau}\frac{\partial v}{\partial t}
\label{gener3}
\end{equation}

Seeking the solution of (\ref{gener3}) as $v=v_0\exp\left(i(kx-\omega t)\right)$ gives $\omega^2+\omega\frac{i}{\tau}-c^2k^2=0$. If $ck<\frac{1}{2\tau}$, $\omega$ does not have a real part and propagating modes. For $ck>\frac{1}{2\tau}$, we find

\begin{equation}
v\propto\exp\left(-\frac{t}{2\tau}\right)\exp(i\omega t)
\label{omega0}
\end{equation}

\noindent where

\begin{equation}
\omega=\sqrt{c^2k^2-\frac{1}{4\tau^2}}
\label{omega}
\end{equation}

An important property is the emergence of the gap in $k$-space: in order for $\omega$ in (\ref{omega}) to be real, $k>k_g$ should hold, where

\begin{equation}
k_g=\frac{1}{2c\tau}
\label{kgap}
\end{equation}

Recently \cite{prl}, detailed evidence for the $k$-gap was presented on the basis of molecular dynamics simulations. According to (\ref{kgap}), the gap in $k$-space increases with temperature because $\tau$ decreases. In agreement with this prediction, Figure 1 shows the $k$-gap emerging in the liquid at high temperature.

\begin{figure}
\begin{center}
{\scalebox{0.8}{\includegraphics{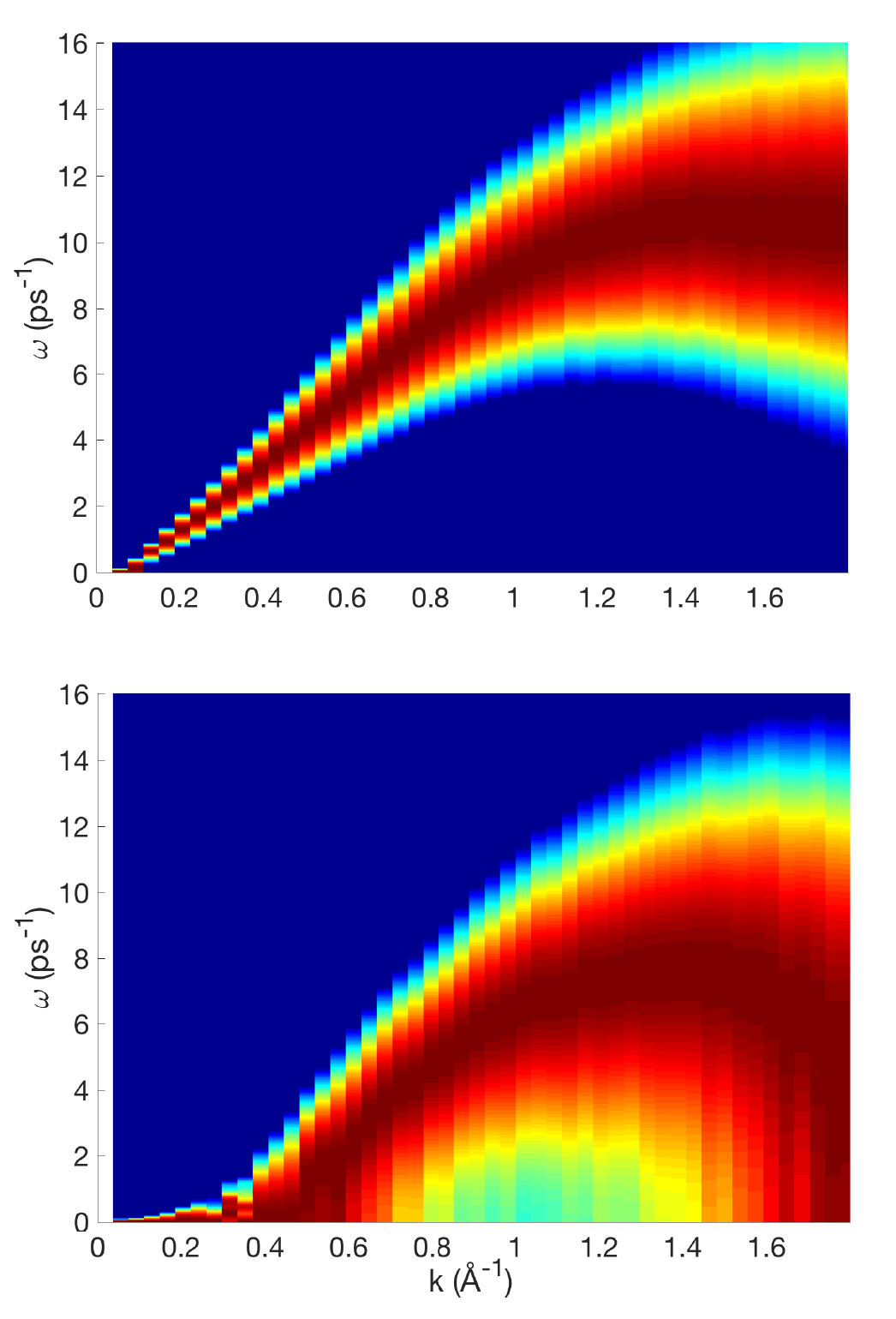}}}
\end{center}
\caption{Intensity maps showing the spectra of transverse currents calculated as the Fourier transform of the real part of transverse current correlation functions. The intensity maps are shown for a model Ar liquid at 205 K (top) and 450 K (bottom) and 10 kbar pressure. The maximal intensity corresponds to the middle points of dark red areas and reduces away from them. The emergence of the gap in $k$-space is seen at high temperature. Adapted from Ref. \cite{prl}.
}
\label{3}
\end{figure}

The gap in $k$-space, or momentum space is interesting. Indeed, the two commonly discussed types of dispersion relations are either gapless as for photons and phonons, $E=p$ ($c=1$), or have the energy (frequency) gap for massive particles, $E=\sqrt{p^2+m^2}$, where the gap is along the Y-axis. On the other hand, (\ref{omega}) implies that the gap is in {\it momentum} space and along the X-axis, similar to the hypothesized tachyon particles with imaginary mass \cite{tachyons}. Figure 2 illustrates this point. This begs the question of what kind of Lagrangian can give (\ref{omega}) and the gap in momentum space.

\begin{figure}
\begin{center}
{\scalebox{0.4}{\includegraphics{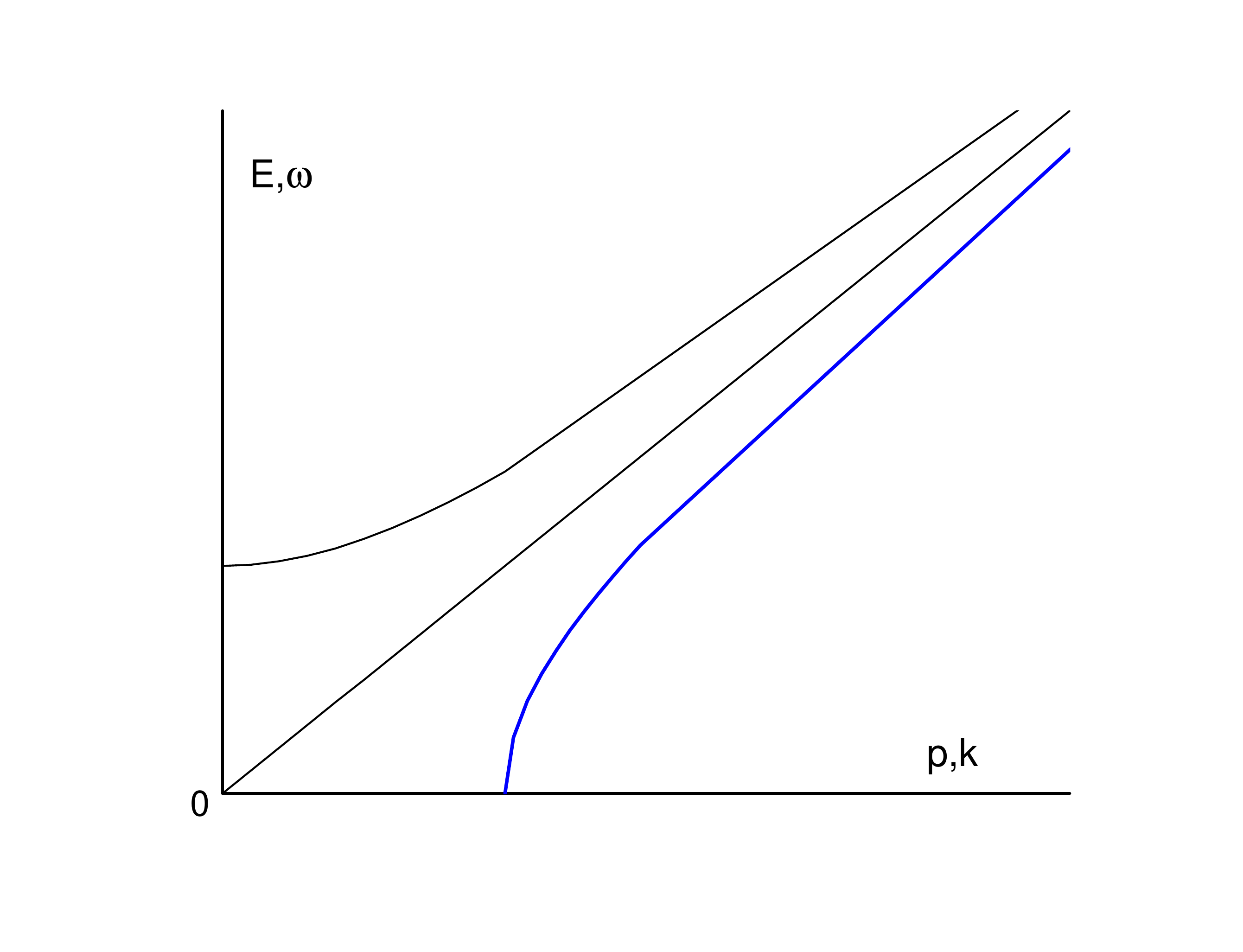}}}
\end{center}
\caption{Possible dispersion relations: dependencies of energy $E$ or frequency $\omega$ on momentum $p$ or wavevector $k$. Top curve shows the dispersion relation for a massive particle with the energy (frequency) gap. Middle curve shows gapless dispersion relation for a massless particle (photon) or a phonon in solids. Bottom curve shows the dispersion relation (\ref{omega}) with the gap in $k$-space. Schematic illustration.}
\label{disp}
\end{figure}

Microscopically, the gap in $k$-space can be related to a finite propagation length of shear waves in a liquid: if $\tau$ is the time during which the shear stress relaxes, $c\tau$ gives the shear wave propagation length \cite{jpcm}. Therefore, the condition $k>k_g=\frac{1}{2c\tau}$ (\ref{kgap}) approximately corresponds to propagating waves with wavelengths shorter than the propagation length.

We now demonstrate that the central equation (\ref{gener3}) giving solid-like shear modes in liquids and their notable gap in $k$-space can be derived by adopting the solid-state description as a starting point. We start with the wave equation describing a non-decayed propagation of transverse waves in the solid:

\begin{equation}
G\frac{\partial^2v}{\partial x^2}=\rho\frac{\partial^2v}{\partial t^2}
\label{2ap}
\end{equation}

We generalize (\ref{2ap}) by substituting $G$ by $M$ (\ref{operator}):

\begin{equation}
G(1-A^{-1})\frac{\partial^2v}{\partial x^2}=\rho\frac{\partial^2v}{\partial t^2}
\label{2ap1}
\end{equation}

We act on both sides of (\ref{2ap1}) with operator $A$ in (\ref{a2}):

\begin{equation}
G\left(A-1\right)\frac{\partial^2v}{\partial x^2}=\rho A\frac{\partial^2v}{\partial t^2}
\label{2ap2}
\end{equation}

According to (\ref{a2}), $A-1$ in (\ref{2ap2}) is $\tau\frac{\partial d}{\partial t}$. Using $G=\rho c^2$ as earlier and rearranging the right side of (\ref{2ap2}) gives

\begin{equation}
c^2\tau\frac{\partial}{\partial t}\frac{\partial^2v}{\partial x^2}=\frac{\partial}{\partial t}\left(\frac{\partial v}{\partial t}+\tau\frac{\partial^2 v}{\partial t^2}\right)
\label{2ap3}
\end{equation}

Integrating over time and setting the integration constant to 0 gives the equation identical to (\ref{gener3}) predicting solid-like shear modes with the gap in $k$-space.

We note that the gap in $k$-space can be inferred from traditional generalized hydrodynamics. Introducing a memory function in the hydrodynamic transverse current correlation function and assuming that the memory function exponentially decays with time $\tau$ gives the current with a resonant frequency similar to (\ref{omega}) \cite{boon}. We derive the gap using Frenkel operators (\ref{a2}) and (\ref{operator}) because this method is simpler and, more importantly, because it enables us to readily apply it to both hydrodynamic and solid-state equations.

We have ascertained the symmetry of liquid description with regard to hydrodynamic and solid-state approaches. We now formulate a Lagrangian theory of liquid dynamics where the dissipative flow and solid-state terms are treated on equal footing, rather than entering sequentially as in the above generalizations of hydrodynamics and elasticity. In addition to liquid theory, an important question from field-theoretical perspective is what Lagrangian gives Eq. (\ref{gener3}) and the associated unusual gap in momentum space (\ref{omega})-(\ref{kgap})? As mentioned earlier, the Lagrangian formulation of liquids involving dissipation was deemed impossible (see, e.g., \cite{leut}).

Our proposed approach is different from the usual way of accounting for dissipation (adding the derivative of the Rayleigh dissipation function to the Euler-Lagrange equation for the non-dissipative Lagrangian) and will result in the Lagrangian with interesting field-theoretical properties. We consider the scalar field $\phi(x,t)$ describing velocities or displacements.

The challenge is to represent the viscous term $\propto\frac{1}{\tau}$ in (\ref{gener3}) in the Lagrangian. The viscous energy can be written as the work $W$ done to move the liquid. If $s$ is the strain, $W\propto Fs$, where $F$ is the viscous force $F\propto\eta\frac{ds}{dt}$. Hence, the Lagrangian should contain the term $s\frac{ds}{dt}$ or, in terms of the field $\phi$, the term

\begin{equation}
L\propto\phi\frac{\partial\phi}{\partial t}
\label{fdf}
\end{equation}

However, the term $\phi\frac{d\phi}{dt}$ disappears from the Euler-Lagrange equation

\begin{equation}
\frac{\partial L}{\partial\phi}=\frac{\partial}{\partial t}\frac{\partial L}{\partial\frac{\partial\phi}{\partial t}}+\frac{\partial}{\partial x}\frac{\partial L}{\partial\frac{\partial\phi}{\partial x}}
\label{lagr}
\end{equation}

\noindent because $\frac{\partial L}{\partial\phi}=\frac{\partial}{\partial t}\frac{\partial L}{\partial\frac{\partial\phi}{\partial t}}=\frac{\partial\phi}{\partial t}$.

To circumvent this problem, we propose to operate in terms of {\it two} fields $\phi_1$ and $\phi_2$, i.e. we invoke two scalar field theory used in a different context \cite{zee} (the same results can be achieved if the complex scalar field theory involving $\phi$ and $\phi^\dagger$ is used). We note that a two-coordinate description of a localised damped harmonic oscillator was discussed earlier \cite{bateman,dekker}.

We construct the dissipative term as a combination of (\ref{fdf}) as

\begin{equation}
L\propto\phi_1\frac{\partial\phi_2}{\partial t}-\phi_2\frac{\partial\phi_1}{\partial t}
\label{twof}
\end{equation}

\noindent In this case, $\frac{\partial L}{\partial\phi_i}\ne \frac{\partial}{\partial t}\frac{\partial L}{\partial\frac{\partial\phi_i}{\partial t}}$, and the dissipative terms remain in the equations of motion for both $\phi_1$ and $\phi_2$.

Using this proposal, we write the Lagrangian as:

\begin{equation}
L=\frac{\partial\phi_1}{\partial t}\frac{\partial\phi_2}{\partial t}-c^2\frac{\partial\phi_1}{\partial x}\frac{\partial\phi_2}{\partial x}+\frac{1}{2\tau}\left(\phi_1\frac{\partial\phi_2}{\partial t}-\phi_2\frac{\partial\phi_1}{\partial t}\right)
\label{l1}
\end{equation}

\noindent where $\frac{\partial\phi_1}{\partial t}\frac{\partial\phi_2}{\partial t}$ and $\frac{\partial\phi_1}{\partial x}\frac{\partial\phi_2}{\partial x}$ are generalizations of the kinetic and potential terms to two-field description.

$\tau\rightarrow\infty$ corresponds to no particle jumps and, therefore, solid dynamics. Setting $\tau\rightarrow\infty$ in (\ref{l1}), we recover the Lagrangian describing waves in solids as expected.

Applying (\ref{lagr}) to the two fields in (\ref{l1}) gives

\begin{equation}
\begin{aligned}
c^2\frac{\partial^2\phi_1}{\partial x^2}=\frac{\partial^2\phi_1}{\partial t^2}+\frac{1}{\tau}\frac{\partial\phi_1}{\partial t}\\
c^2\frac{\partial^2\phi_2}{\partial x^2}=\frac{\partial^2\phi_2}{\partial t^2}-\frac{1}{\tau}\frac{\partial\phi_2}{\partial t}
\end{aligned}
\label{2eq}
\end{equation}

The first equation in (\ref{2eq}) is identical to (\ref{gener3}), resulting in the solution (\ref{omega0}-\ref{omega}): $\phi_1=\phi_0\exp\left(-\frac{t}{2\tau}\right)\cos(kx-\omega t)$ with $\omega=\sqrt{c^2k^2-\frac{1}{4\tau^2}}$. The second equation gives the solution growing with time: $\phi_2=\phi_0\exp\left(\frac{t}{2\tau}\right)\cos(kx-\omega t)$ with the same $\omega$ (for simplicity, we assume zero phase shifts in $\phi_1$ and $\phi_2$). This solution can be discounted as non-physical or viewed as the wave propagating back in time and space with respect to $\phi_1$ because
\begin{equation}
\phi_2(x,t)=\phi_1(-x,-t)
\end{equation}

This implies that a Lagrangian formulation of a non-reversible dissipative process necessitates two waves moving in the opposite space-time directions, resulting in this sense in the reversibility of the Lagrangian description.

The total energy of the system does not have exponential terms $\exp\left(\pm\frac{t}{2\tau}\right)$ due to their cancellation. Indeed, the Hamiltonian is $H=\pi_1\frac{\partial\phi_1}{\partial t}+\pi_2\frac{\partial\phi_2}{\partial t}-L$, where $\pi_1=\frac{\partial\phi_2}{\partial t}-\frac{\phi_2}{2\tau}$ and $\pi_2=\frac{\partial\phi_1}{\partial t}+\frac{\phi_1}{2\tau}$ from (\ref{l1}). This gives $H=\frac{\partial\phi_1}{\partial t}\frac{\partial\phi_2}{\partial t}+c^2\frac{\partial\phi_1}{\partial x}\frac{\partial\phi_2}{\partial x}$. Using the solutions $\phi_1$ and $\phi_2$ above gives the system energy $E=\phi_0^2\left(2c^2k^2\sin^2(kx-\omega t)-\frac{1}{4\tau^2}\right)$. Averaged over time, $E=\phi_0^2\left(c^2k^2-\frac{1}{4\tau^2}\right)=\phi_0^2\omega^2$ and is constant and positive.

We now make several observations regarding the field-theoretical description of liquid dynamics ({\ref{l1}). The first two terms in ({\ref{l1}) have the form of relativistic scalar field theory, but the last term in brackets has the form of non-relativistic approximation to the relativistic term (see, e.g., \cite{zee}). Perhaps for this reason, there has been no rationale to write (\ref{l1}) in field theory, and this Lagrangian was not discussed. In our approach, the last term in (\ref{l1}) represents dissipative hydrodynamic motion. From the point of view of field theory, this term represents a way to treat strongly anharmonic self-interaction of the field. Indeed, if this interaction has a double-well (or multi-well) form, the field can move from one minimum to another in addition to oscillating in a single well \cite{prd}. This motion is analogous to diffusive particle jumps in the liquid responsible for the viscous $\propto\frac{1}{\tau}$ term in (\ref{gener3}). Therefore, the dissipative $\propto\frac{1}{\tau}$ term in (\ref{l1}) and (\ref{2eq}) describes the hopping motion of the field between different wells with frequency $\frac{1}{\tau}$. We refer to this term as dissipative, although we note that no energy dissipation takes place in the system as discussed above. Rather, the dissipation concerns the propagation of plane waves in the anharmonic field of Lagrangian (\ref{l1}). The dissipation varies as $\propto\frac{1}{\tau}$: large $\tau$ corresponds to rare transitions of the field between different potential minima and non-dissipative wave propagation due to the first two terms in (\ref{l1}).

From the point of view of field theory, it is interesting to see how the gap in $k$ space in (\ref{l1}) can vary and to elaborate on general effects of the dissipative term (\ref{twof}). We start with adding the mass term to (\ref{l1}), $-\Omega^2\phi_1\phi_2$:

\begin{equation}
\begin{split}
&L=\frac{\partial\phi_1}{\partial t}\frac{\partial\phi_2}{\partial t}-c^2\frac{\partial\phi_1}{\partial x}\frac{\partial\phi_2}{\partial x}+\\
&\frac{1}{2\tau}\left(\phi_1\frac{\partial\phi_2}{\partial t}-\phi_2\frac{\partial\phi_1}{\partial t}\right)-\Omega^2\phi_1\phi_2
\end{split}
\label{lagr1}
\end{equation}
\noindent where $\Omega$ is bare mass.

Seeking the solution in the form of a plane wave as before we find the real part of $\omega$ corresponding to propagating waves for both $\phi_1$ and $\phi_2$ as

\begin{equation}
\omega=\sqrt{c^2k^2+\Omega^2-\frac{1}{4\tau^2}}
\label{newomega}
\end{equation}

%In the non-dissipative case $\tau\rightarrow\infty$, (\ref{newomega}) has the energy gap $\Omega$ as expected: removing the viscous terms in (\ref{lagr1}) by setting $\tau\rightarrow\infty$ gives the Klein-Gordon equation with the usual energy (mass) gap $\Omega$.

$\omega$ in (\ref{newomega}) behaves differently depending on the sign of $\Omega^2-\frac{1}{4\tau^2}$. If $\Omega>\frac{1}{2\tau}$, (\ref{newomega}) gives the mass (energy) gap as

\begin{equation}
\omega(k=0)=\sqrt{\Omega^2-\frac{1}{4\tau^2}}
\label{mass-em}
\end{equation}

We observe that the dissipative term reduces the mass (energy gap) from its bare value $\Omega$.

When $\Omega<\frac{1}{2\tau}$, (\ref{newomega}) predicts the gap in $k$-space. Indeed, under this condition the expression under the square root in (\ref{newomega}) is negative unless $k>k_g$, where

\begin{equation}
k_g=\sqrt{\frac{1}{4c^2\tau^2}-\frac{\Omega^2}{c^2}}
\label{newgap}
\end{equation}

Comparing with (\ref{kgap}) we see that the bare mass reduces the gap in $k$-space.

The mass gap and $k$-gap both close when

\begin{equation}
\Omega=\frac{1}{2\tau}
\label{compete1}
\end{equation}

\noindent i.e. when the bare mass $\Omega$ becomes close to the field hopping frequency. In this case, (\ref{newomega}) gives the photon-like dispersion relation $\omega=ck$, corresponding to the first two terms of (\ref{lagr1}) only.

We thus find that the mass and the dissipative terms compete by promoting gaps in energy and $k$-space, respectively. We also find that when both gaps close, the dissipative term annihilates the bare mass term in the dispersion relation. We note that the mass gap can also {\it emerge}. The hopping frequency $\frac{1}{2\tau}$ can be initially large at high energy or temperature and close to the bare mass $\Omega$, giving zero mass gap in (\ref{mass-em}). As the system cools down and $\frac{1}{2\tau}$ decreases, the mass gap becomes close to the bare mass.

(\ref{l1}) can be written in a more symmetrical form by adding the term with spatial derivatives:

\begin{equation}
\begin{split}
&L=\frac{\partial\phi_1}{\partial t}\frac{\partial\phi_2}{\partial t}-c^2\frac{\partial\phi_1}{\partial x}\frac{\partial\phi_2}{\partial x}+\\
&\frac{1}{2\tau}\left(\phi_1\frac{\partial\phi_2}{\partial t}-\phi_2\frac{\partial\phi_1}{\partial t}\right)-
\frac{l}{2\tau^2}\left(\phi_1\frac{\partial\phi_2}{\partial x}-\phi_2\frac{\partial\phi_1}{\partial x}\right)
\end{split}
\label{symL}
\end{equation}
\noindent where $l$ is a certain length scale to be discussed later.

Applying (\ref{lagr}) to (\ref{symL}) and seeking the plane-wave solution gives the real part of $\omega$ as

\begin{equation}
\omega=\frac{1}{\sqrt{2}}\sqrt{c^2k^2-\frac{1}{4\tau^2}+\sqrt{\left(c^2k^2-\frac{1}{4\tau^2}\right)^2+\left(\frac{lk}{\tau^2}\right)^2}}
\label{omega1}
\end{equation}

\noindent which reduces to (\ref{omega}) if $l=0$.

The expression under the square root in (\ref{omega1}) is always positive and no gap in $k$-space emerges ($k=0$ gives $\omega=0$). Hence the introduction of symmetry with respect to $t$ and $x$ in the Lagrangian removes the $k$-gap. We also observe that setting $l$ in (\ref{omega1}) to $c\tau$, close to the wave propagation length, gives the linear dispersion law $\omega=ck$. This may be interpreted as a result of limiting the length scale to the wave propagation length, i.e. considering a non-dissipative case.

The effect of mass term on (\ref{symL}) can be seen by adding $-\Omega^2\phi_1\phi_2$ and, for definitiveness, setting $l=c\tau$ as above. This gives the real part of $\omega$ as

%\begin{equation}
%\begin{split}
%&L=\frac{\partial\phi_1}{\partial t}\frac{\partial\phi_2}{\partial t}-c^2\frac{\partial\phi_1}{\partial x}\frac{\partial\phi_2}{\partial x}+
%\frac{1}{2\tau}\left(\phi_1\frac{\partial\phi_2}{\partial t}-\phi_2\frac{\partial\phi_1}{\partial t}\right)-\\
%&\frac{c}{2\tau}\left(\phi_1\frac{\partial\phi_2}{\partial x}-\phi_2\frac{\partial\phi_1}{\partial x}\right)-\Omega^2\phi_1\phi_2
%\end{split}
%\label{symL2}
%\end{equation}

\begin{equation}
\begin{split}
&\omega=\frac{1}{\sqrt{2}}\times\\
&\sqrt{c^2k^2+\Omega^2-\frac{1}{4\tau^2}+\sqrt{\left(c^2k^2+\Omega^2-\frac{1}{4\tau^2}\right)^2+\left(\frac{ck}{\tau}\right)^2}}
\end{split}
\label{omega2}
\end{equation}

\noindent which reduces to $\omega=ck$ if $\Omega=0$.

The energy gap in (\ref{omega2}) is the same as (\ref{mass-em}): $\omega(k=0)=\sqrt{\Omega^2-\frac{1}{4\tau^2}}$ and closes when $\Omega=\frac{1}{2\tau}$. Similarly to (\ref{omega1}), no gap in $k$-space emerges. On further increase of the dissipative term $\frac{1}{\tau}$, $\omega$ remains gapless and tends to 0.

To summarize, we demonstrated the equivalence of hydrodynamic and solid-state approaches to liquids and used solid-like shear modes in liquids with the gap in $k$ space as a case study. This symmetry of liquid description is important from the point of view of general outlook on the liquid state of matter. From a more practical perspective, the applicability of solid-like description to liquids means that we can appropriately use the wealth of well-understood effects from solid state physics including collective modes. This importantly applies to the range of large $\omega$ and $k$ not accessible to generalized hydrodynamics (e.g. when a reliable extrapolation is unknown or does not exist). The proposed Lagrangian provides new mechanisms for the interplay between the dissipative and mass terms. A wide range of field-theoretical methods can be now applied to the Lagrangian to study its properties further and understand its applicability in other areas (e.g., neutrinos).

I am grateful to S. Ramgoolam, A. Polnarev, V. Brazhkin and A. Zaccone for discussions and to the Royal Society and EPSRC for support.

\end{document}